\documentclass[aps,pre,onecolumn,12pt,notitlepage,
 showpacs,floatfix,nofootinbib,
 double-space
 superscriptaddress
 ]{revtex4-2}
\usepackage{natbib}
\usepackage[export]{adjustbox}
\usepackage{graphicx}
\usepackage{subcaption}
\usepackage{bbold}
\usepackage{amssymb}
\usepackage{amsfonts}
\usepackage{amsmath}
\usepackage{amsthm}
\usepackage{epsfig}
\usepackage{float}
\usepackage{multirow}
\usepackage[usenames,dvipsnames]{color}
\usepackage[latin1]{inputenc}
\usepackage{xr}
\usepackage{enumitem}
\usepackage{microtype}
\usepackage{graphics,graphicx,xcolor}
\usepackage{comment}
\usepackage{cancel}

\graphicspath{{newfigs/}}

\begin{document}

\title{Speciation, extinction and explosions induced by sudden environmental changes}

\author{Larissa G. Landucci}

\author{Marcus A. M. de Aguiar}

\affiliation{Instituto de F\'isica ``Gleb Wataghi", Universidade Estadual de Campinas, Unicamp 13083-970, Campinas, SP, Brazil}
	
\begin{abstract}

We study the evolution of sexually reproducing populations subjected to sudden environmental changes. Using an individual based model we consider scenarios where abrupt changes can affect the carrying capacity, the mating range or the strength of assortativity, that controls the minimum similarity between individuals that allows mating to occur. We show that a boost in carrying capacity can either lead to new speciation events, thereby increasing biodiversity, or to widespread extinctions, causing the system to become dominated by a single species. A similar effect is observed with respect to the mating range. We show that the key factor determining the outcome of the process is genome length, i.e., the number of genes controlling reproduction and assortativity. We also show that when mating becomes more restrictive, increasing the degree of assortativity, a burst of speciation occurs. The explosion lasts only for a few generations, converging to values larger than before the change, but not as high as at the peak. 

\end{abstract}

\maketitle

\section{INTRODUCTION}
\label{sec:intro}

The origin and maintenance of biodiversity depends on several factors, such as selection, mutations, ecological interactions and geographic features \cite{coyne2004speciation,nosil2012ecological}. Moreover, all these components change over time, adding an extra layer of complexity to the processes of speciation and extinction that ultimately shape biodiversity. Changes in environmental conditions can be the consequence of natural phenomena or the result of human activity \cite{dornelas2023looking}. The current increase in temperature \cite{waldock2018temperature}, deforestation \cite{hagen2012biodiversity}, glacial eras, earthquakes and separation of continents are some examples \cite{benton2016origins}. Although a lot of attention has been given recently to climate change, and how that might cause drastic decrease in biodiversity \cite{thomas2004extinction, urban2015accelerating, maclean2011recent, bellard2012impacts}, little has been done to understand how changes in the environment could have contributed to shape the diversity we see today. Rapid extinctions and formation of species, for example, have been connected to environmental cataclysmic events \cite{vrba1985environment, vrba1993turnover, desbois2025elisabeth}. 

An important framework to describe the effects of sudden environmental changes in evolutionary history is provided by neutral theories. If changes are fast compared to the time scale of natural adaptation, neutral theories offer a simplified way to understand their effects on population structure. In this paper we address the question of how sudden changes affect species diversity using a simple evolutionary model. We extend the individual based model proposed by de Aguiar et al. \cite{de2009global}, designed to study the evolution of sexually reproducing populations under constant conditions, to describe environmental changes. The model successfully predicted empirical patterns of abundance distributions and species-area relationship \cite{de2009global}. More recently, it has been also shown that model parameters can be scaled to realistic values to describe speciation of living organisms \cite{nelson2024neutral}. 

Several extensions of the model presented in \cite{de2009global} have been proposed in the past years, including applications to ring species, mito-nuclear co-evolution, island biogeography, antagonistic interactions and environmental selection \cite{baptestini2013role, martins2013evolution, schneider2014toward, schneider2016diploid, de2017speciation, marquioni2025transition, schneider2016mutation, baptestini2013conditions, oliveira2014modelo, botelho2022extinction, schneider2014topopatric, nelson2024neutral, marquioni2025transition, princepe2022diversity, princepe2022mito}. All these applications, however, consider that evolution occurs under a constant external environment, an assumption implicit in the fixed parameters of the model. Here, instead, we propose to model sudden events by allowing the values of the model parameters to change during the evolutionary time. We are interested, in particular, to model pulses events, where ecological parameters change abruptly \cite{jentsch2019theory}.

Most of our simulations will involve change in parameters that control carrying capacity (the maximum number of individuals supported by the environment); mating range (the area around the individual where it searches for a mate); and degree of choosiness, or assortativity (minimum similarity between individuals that allow them to mate or result in viable offspring). Mating range and carrying capacity depend on several factors, such as temperature and humidity \cite{martire2015carrying}. Choosiness, on the other hand, can be affected by species specific factors. Cychlid fish, for example, choose mates by visual cues, which depends on the transparency of water to be effective. Turbid water hinders identification of similar individuals and lead to more relaxed assortativity \cite{seehausen1997cichlid,seehausen2008speciation}. Clear waters, on the other hand, enhance visual recognition and leads to higher degree of choosiness \cite{seehausen2008speciation}.

From the evolutionary point of view, increasing the carrying capacity leads to larger populations and, consequently, larger number of mutations per generation. We will show that, depending on the number of genes controlling reproduction, a population with given number of species might break up into even more species, increasing diversity, or undergo extinctions, resulting in a single dominant species and drastically reducing the biodiversity. A similar phenomenon is observed with respect to the mating range. Perhaps more interesting, we will show that a sudden increase in the degree of choosiness, such as turning water clearer in the case of cyclids, can cause a transient explosion in the number of species, that subsequently converge to values larger than before the event. Different combinations of carrying capacity and mating range also affect the explosion's peak when the degree of assortativity is changed.

The remainder of this paper is organized as follows: in section \ref{sec:model} we describe the individual based model used in the simulations. In the model, populations are distributed uniformly over a living area, and genomes are characterized by a string of biallelic genes. Both finite and infinite strings are considered. Section \ref{sec:results} contains the result of our simulations and in section \ref{sec::discussion} we discuss our findings.

\section{Evolutionary Model}
\label{sec:model}

We consider a finite population distributed on a square area of fixed size with either periodic or reflective boundary conditions. To simplify the simulations, space is modeled as  $L \times L$ lattice with unitary spacing, leading to $L^2$ points representing the possible positions of individuals. More than one individual can occupy the same lattice point, although this is unlikely to happen, as typical simulations have low density of individuals. The environment has ecological carrying capacity $M_0$, representing the number of individuals that can be supported by local resources. We assume that basal resources are renewed in each generation and are also distributed uniformly. In the first generation the population size is set to $M_0$.  

Individuals are characterized by their spatial location $(x^\alpha,y^\alpha)$ and by a haploid genome $S^{\alpha}$, represented by a sequence of $B$ bi-allelic genes
\begin{equation}\label{genome}
      {S}^{\alpha}=\{{S}_{1}^{\alpha},{S}_{2}^{\alpha},...,{S}_{B}^{\alpha}\}, 
\end{equation}
where $S_k^\alpha$ can take the values $\pm 1$. We use Greek letters to label individuals and Latin letters for genome loci.

The genetic similarity between two individuals $\alpha$ and $\beta$ is defined by
\begin{equation}
     q^{\alpha\beta}= \frac{1}{B} \sum_{k=1}^B S_k^\alpha S_k^\beta
     \label{eqsim}
\end{equation}
and goes from $+1$, when the genomes are identical ($S_k^\alpha = S_k^\beta$), to $-1$, when genomes differ at all loci ($S_k^\alpha =- S_k^\beta$). Equivalently, we define relative the genetic distance as
\begin{equation}
     d^{\alpha\beta}=\frac{1}{2}\left( 1- q^{\alpha\beta} \right),
     \label{eqdis}
\end{equation}
representing the fraction of loci where the genomes differ.

Initially, the individuals are genetically identical, with $S_k^\alpha=1$ for $k=1,2,\dots,B$ and $\alpha=1,2,\dots M$, and uniformly random distributed in space. The model describes the evolution of the population simulating sexual reproduction with recombination and mutation. Individuals are assumed to be hermaphrodites, with no separation into males and females.  

Following de Aguiar et al \cite{de2009global} we allow mating between two individuals only if the following two conditions are met: (i) they must be sufficiently close in space, so that their Euclidean distance satisfies $r^{\alpha\beta} \leq R$ and, (ii) they must be sufficiently similar (or genetically compatible) so that $q^{\alpha\beta} \geq q_{min}$. The circular area of radius $R$ around an individual is called its {\it mating neighborhood}, and $q_{min}$ is the {\it similarity threshold}. In terms of genetic distance, the latter condition reads $d^{\alpha\beta} \leq g$ where $g=(1-q_{min})/2$. Less restrictive criteria for genetic compatibility were explored in ref. \cite{lizarraga2024assortativity}.

Generations are non-overlapping, so that the current population is completely replaced by the new one. 
Each individual in the current generation is selected once to reproduce and is termed the {\it focal} parent. A genetically compatible mating partner is then selected from its mating neighborhood, as described below for finite and infinite genomes.  If no compatible partner is found, the individual dies without leaving an offspring.
In contrast, if one or more genetically compatible partners are found, one is selected at random and one offspring can be generated. If the current population size $M$ is larger than the carrying capacity $M_0$, there's a probability of $0.15(1-M_0/M)$ that the individual dies without leaving descendants. However, if  $M < M_0$ and the local density is low, there is also a probability $0.5(1-M/M_0)$ that the individual will leave two offspring with the same partner. For this we require that the density of individuals in the mating neighbourhood of the focal does not exceed $a \frac{M_0}{\pi R^2}$, where $a=1.2$ (for reflective boundaries) and $a=1$ (for periodic boundaries), and $\frac{M_0}{\pi R^2}$ is the expected density at carrying capacity. This condition prevents local overcrowding. The values of the parameter $a$ and the coefficients multiplying the probabilities of dying or having two offspring were adjusted to allow for small fluctuations of $M$ around $M_0$. The offspring are placed at random positions in the focal parent's mating neighborhood.

\subsection{Finite genomes}

If genome size $B$ is finite, mating compatibility between individuals $\alpha$ and $\beta$ require that $q^{\alpha\beta} \geq q_{min}$, or, equivalently, that the fraction of loci bearing different alleles is less or equal to $g$. In that case the offspring inherits, gene by gene, the allele from one of its parents, focal or mating partner, with equal probability. The allele is than subjected to mutation, from 1 to -1 or vice-versa, with probability $\mu$. 

\subsection{Infinite genomes}

For infinite genomes, given the impossibility of an explicit treatment, we directly update the matrix $q^{\alpha\beta}$ of genetic similarities. Following  \cite{higgs1991stochastic, de2017speciation} we let $\mu$ be the mutation {\it rate} and consider first asexual reproduction. In this case each individual $\alpha$ has a single parent $P(\alpha)$ in the previous generation. The probability that $\alpha$ keeps the allele of its parent is $(1+e^{-2\mu})/2 \approx 1-\mu$, which is the probability that it does not mutate in a time interval of $1$ generation. Similarly, the probability that it mutates is $(1-e^{-2\mu})/2 \approx \mu$. The expected value of the allele is, therefore 
\begin{displaymath}
    E(S^\alpha_k) = S^{P(\alpha)}_k \; \frac{1+e^{-2\mu}}{2} - S^{P(\alpha)}_k \; \frac{1-e^{-2\mu}}{2} = e^{-2\mu} S^{P(\alpha)}_k. 
\end{displaymath}

For the case of sexual reproduction, each parent has probability $1/2$ to pass its gene to the offspring leading to
\begin{equation}
     E(S^\alpha_k) = \frac{e^{-2\mu}}{2} \left( S^{P_1(\alpha)}_k + S^{P_2(\alpha)}_k \right).
     \label{avsex}
\end{equation}
If genes are independent we obtain (see Eq.(\ref{eqsim})),
\begin{equation}\label{expdist}
     E(q^{\alpha\beta}) =\frac{e^{-4\mu}}{4}(q^{P_{1}^\alpha P_{1}^\beta}+q^{P_{1}^\alpha P_{2}^\beta}+q^{P_{2}^\alpha P_{1}^\beta}+q^{P_{2}^\alpha P_{2}^\beta}),
\end{equation}
where $P_{1}^\alpha$ and $P_{2}^\alpha$ are parents of $\alpha$ and $P_{1}^\beta$ and $P_{2}^\beta$ are parents of $\beta$. We also define $q^{\alpha \alpha}\equiv{1}$. Since the number of genes is infinite, the average converges to the exact result. Dropping the average symbol, Eq.(\ref{expdist}) can be used to update the similarity matrix generation by generation.

\subsection{Species}

In the model, species are identified as sets of individuals connected by potential gene flow, and can be computed as follows. We construct a network where individuals are nodes and edges connect nodes that are reproductively compatible. The corresponding adjacency matrix is defined such that $A_{ij}=1$ if $q^{ij} \geq q_{min}$, and $A_{ij}=0$ otherwise. Species are defined as connected components of this network, which correspond to groups of individuals connected by potential gene flow and fully disconnected from other components. 

\subsection{Parameters and initial conditions}

Before we show the results of numerical simulations we highlight some of its key parameters and their role in promoting or hinder speciation. The main ingredients of the model can be summarized as follows: (i) genetic information is described by a single (haploid) binary chain representing independent bi-allelic genes; (ii) mating between two individuals is restricted by spatial proximity and genetic similarity and; (iii) when reproduction occurs, the offspring inherits the alleles of each parent with equal probability followed by mutation.  This simplified description involves 6 parameters: genome size ($B$); area of habitat ($L^2$); radius of mating neighborhood ($R$); threshold similarity ($q_{min}$); mutation probability ($\mu$) and; carrying capacity, ($M_0$).

At the start of the simulations we assume that all individuals are genetically identical and, therefore, genetically compatible. Mutations introduce variations that can only be transmitted locally, to an offspring inside the mating neighborhood, at the next generation. Accordingly, genetic changes travel slowly through space and distant regions develop nearly independent modifications in the genomes. Depending on parameter values this can lead to the formation of groups of individuals that are genetically incompatible with each other, i.e., to species.  Genetically similar individuals tend to inhabit the same region of space, creating a strong correlation between genetic and physical space.

In the next section we will show results where parameters $q_{min}$, $R$ and $M_0$ change during the evolution. 

\section{Results}
\label{sec:results}

The following results were obtained using $\mu$ = 0.0005 and $L=100$, leading to a lattice ${100}\times{100}$. Average values correspond to ten repetitions of the simulations. We compare results of finite genomes with length $B=1500$ and infinite genomes, both with reflective and periodic boundaries. Changes in the parameters were performed at $t=800$, which is enough for the system to have reached equilibrium with the initial parameter values. Simulations were run for a total of $T=1500$ generations. We refer to the four combinations of genome length and boundary conditions as FR, FP, IR and IP, where F and I stand for Finite and Infinite genome length and R and P to Reflective and Periodic boundary conditions. Parameter values are chosen to generate sufficient variation in a relatively small number of generations and make computational time manageable. 

\begin{figure}[h]
    \begin{subfigure}{0.4\textwidth}
    \centering
    \includegraphics[width=\linewidth]{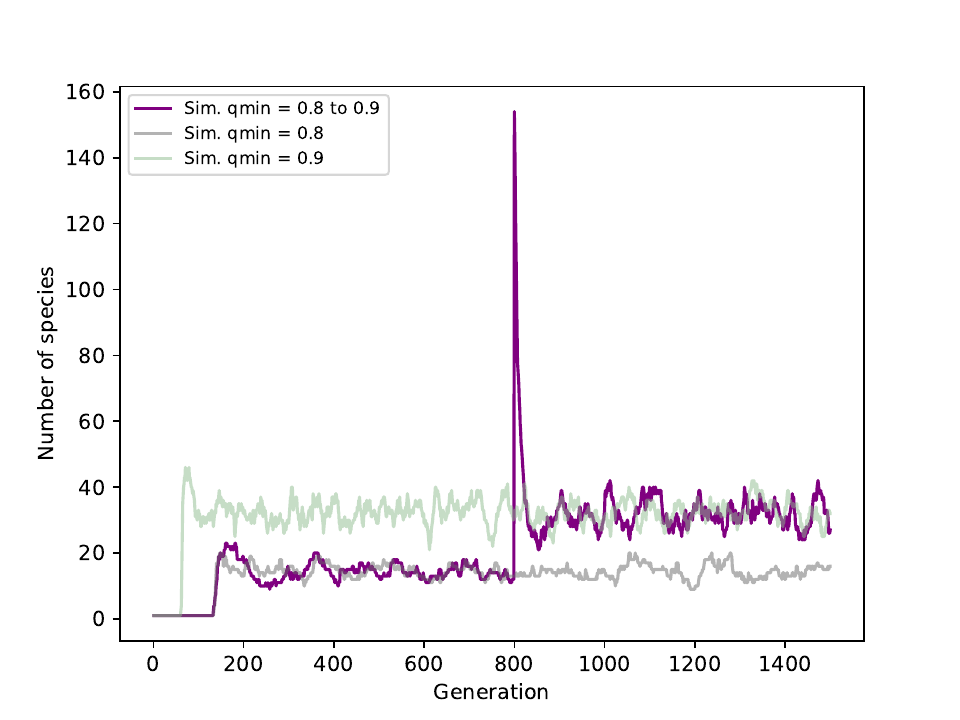}
    \caption{}
    \label{Nspcgen_qminhigher}
    \end{subfigure}
    \begin{subfigure}{0.4\textwidth}
    \centering
    \includegraphics[width=\linewidth]{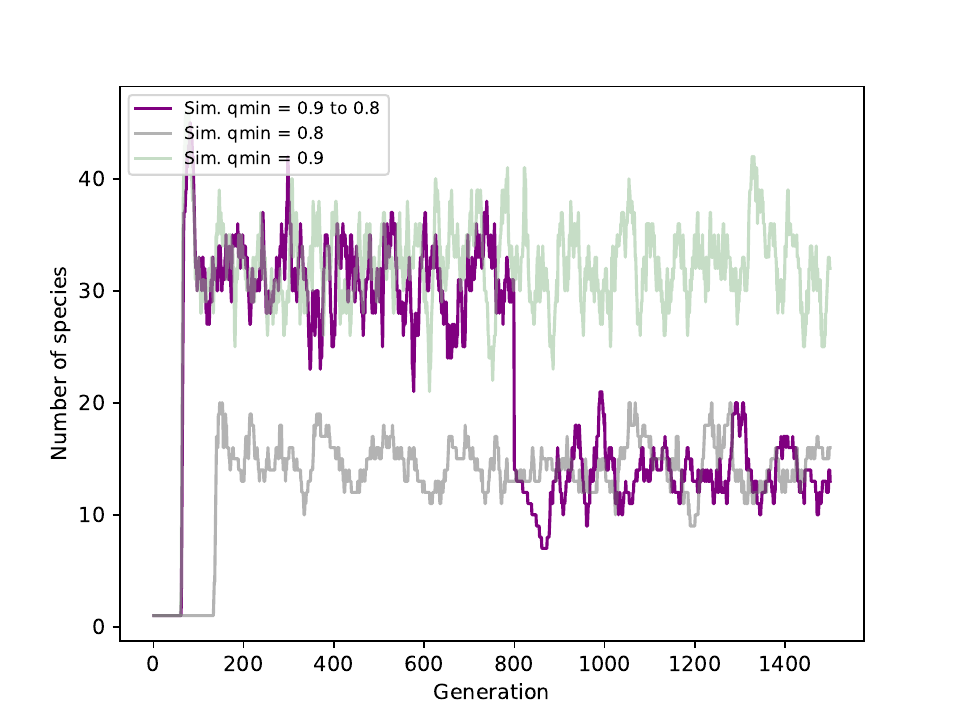}
    \caption{}
    \label{Nspcgen_qminlower}
    \end{subfigure}   
\caption{Number of species as a function of time for $M_0=1300$, $R=5.0$ and IP scenario (infinite genome length and periodic boundary conditions). The similarity threshold $q_{min}$ changed at $t=800$ from (a) $0.8$ to $0.9$ and (b) $0.9$ to $0.8$ (purple curves). Gray and green curves correspond to evolution with constant $q_{min}=0.8$ and $0.9$, respectively.}
\label{Nspcgen}
\end{figure}

As mentioned in the introduction, the change that causes the most interesting variations in system dynamics is in the degree of assortativity, related to the genetic similarity threshold $q_{min}$. Figure \ref{Nspcgen} shows how the number of species change when $q_{min}$ changes from $0.8$ to $0.9$, representing the sudden introduction of a more restrictive genetic barrier, and from $q_{min}=0.9$ to $0.8$, corresponding to more permissive mating, or a less restrictive genetic barrier. The simulations shown were performed for infinite genomes and periodic boundary conditions (IP). When mating becomes more restrictive, species break into smaller groups, leading to a large peak in diversity. The peak, however, is short lived, lasting between 30 and 50 generations, and is accompanied by a reduction in population size, leading to low abundances in the newly formed species. After this transient period the number of species stabilizes at the expected value corresponding to the new set of parameters. The transition from more to less restrictive mating (panel (b)) is less dramatic, although a small decrease in the number of species, to values below the average, is observed right after the transition.

\begin{figure}[h]
    \begin{subfigure}{0.4\textwidth}
    \centering
    \includegraphics[width=\linewidth]{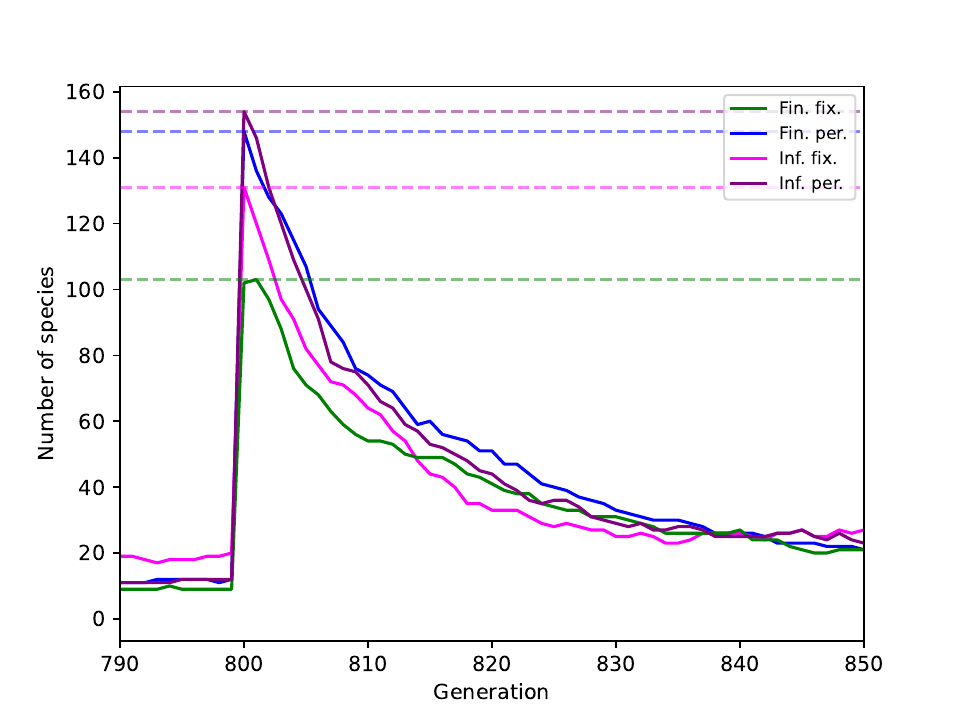} 
    \caption{}
    \label{PeaksNspc}
    \end{subfigure}
    \begin{subfigure}{0.4\textwidth}
    \centering
        \includegraphics[width=\linewidth]{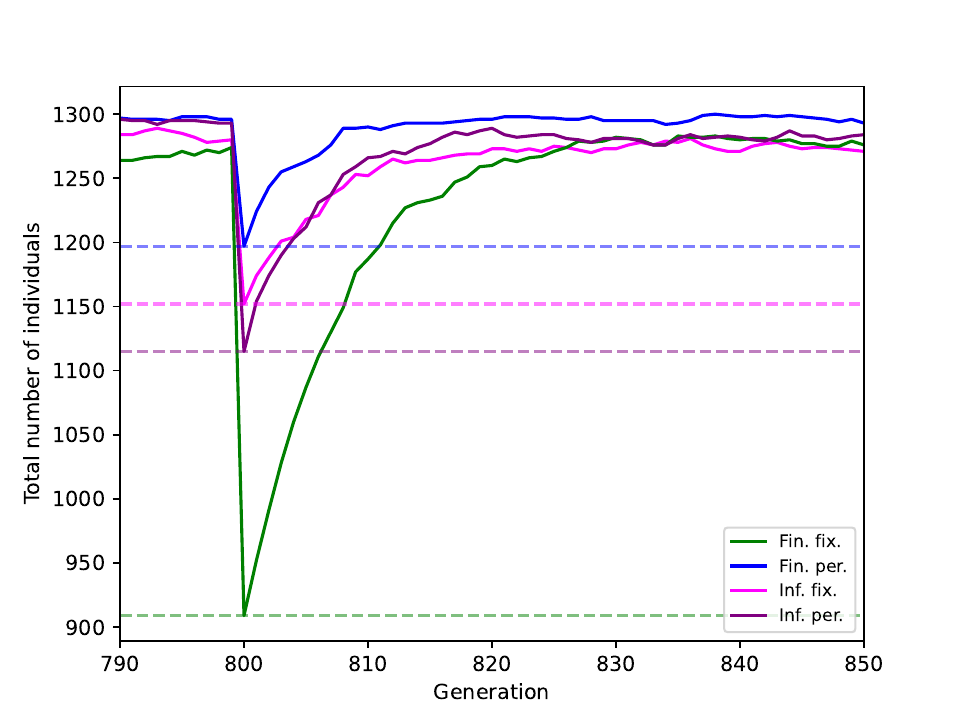}
    \caption{}
    \label{PeaksNind}
    \end{subfigure}
     \begin{subfigure}{0.4\textwidth}
    \centering
    \includegraphics[width=\linewidth]{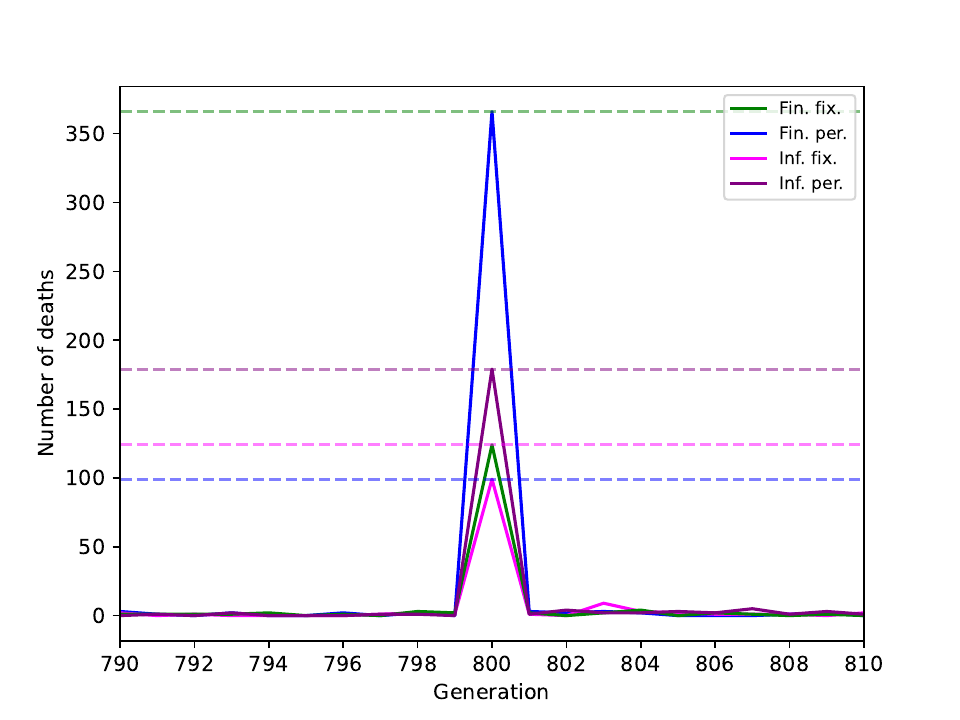}
    \caption{}
    \label{PeaksNdeaths}
    \end{subfigure}
    \begin{subfigure}{0.4\textwidth}
    \centering
    \includegraphics[width=\linewidth]{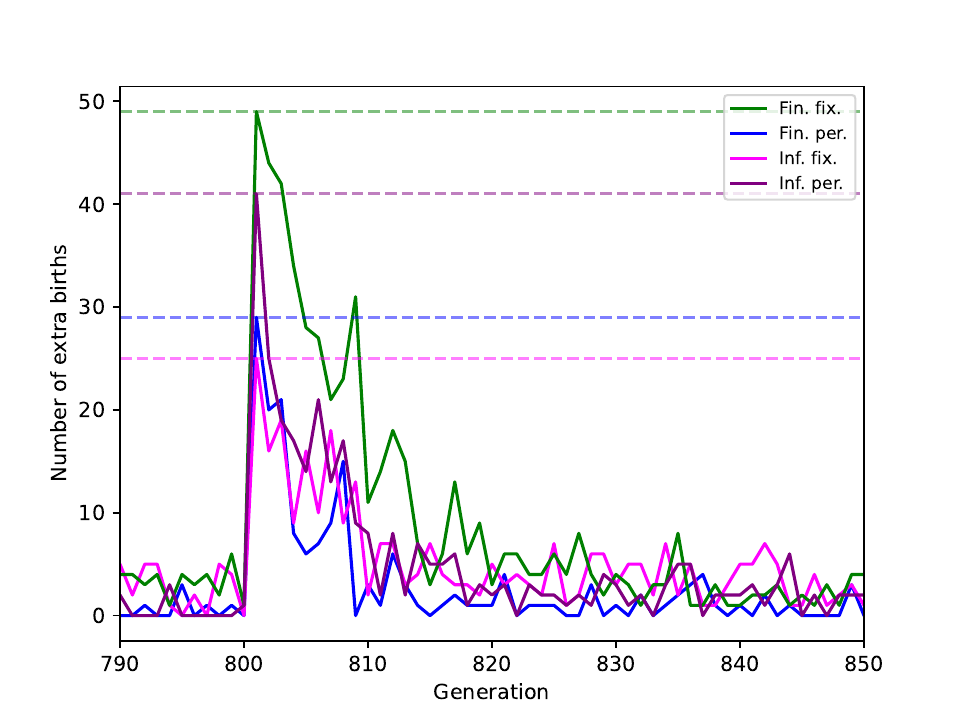}
    \caption{}
    \label{PeaksNbirths}
    \end{subfigure}   
\caption{Details of the explosive transition from $q_{min}=0.8$ to $0.9$. (a) number of species; (b) population size; (c) number of individuals that did not leave descendants and; (d) number of extra births. The purple curves correspond to infinite genome length and periodic boundary conditions (IP), as shown in Fig.\ref{Nspcgen}. Magenta curves show results for infinite genome and reflective boundaries (IR), whereas green and blue lines correspond to finite genomes and reflective (FR) or periodic boundaries (FP), respectively.}
\label{Peaks}
\end{figure}

Figure \ref{Peaks} shows in more the detail the species explosion at the transition from $q_{min}=0.8$ to $0.9$. The sudden peak in the number of species, Fig.\ref{Peaks}(a), drops smoothly to its equilibrium after about 50 generations for all combinations of boundary conditions and genome lengths. The height of the peak, however, depends on both. The largest species diversity at the transition is achieved by the IP setting and the smallest by the FR. The drop in population size, Fig.\ref{Peaks}(b), on the other hand, is largest for FR and smallest for FP. In all cases, population size recovers faster (about 20 generations) than number of species. Fig.\ref{Peaks}(c) shows the number of individuals per generation that could not find a compatible mating partner and died without leaving offspring. This number is large only at $t=800$, when the change in $q_{min}$ happens. For $t>800$ the population becomes smaller, but the remaining individuals do have compatible mates in their neighborhoods, which ensures the subsequent growth of the population. This is also shown in Fig.\ref{Peaks}(d), displaying how many {\it extra} offspring were generated, beyond the usual one offspring per mating event.

\begin{figure}[h]
     \begin{subfigure}{0.4\textwidth}
    \centering
    \includegraphics[width=\linewidth]{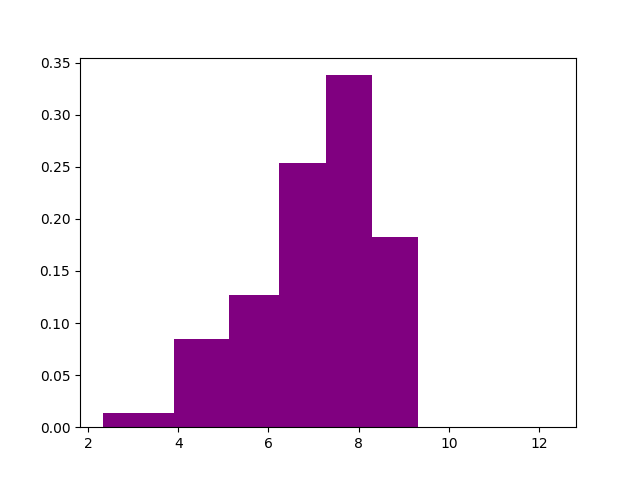} 
    \caption{}
    \label{HistSize799}
    \end{subfigure}
    \begin{subfigure}{0.4\textwidth}
    \centering
        \includegraphics[width=\linewidth]{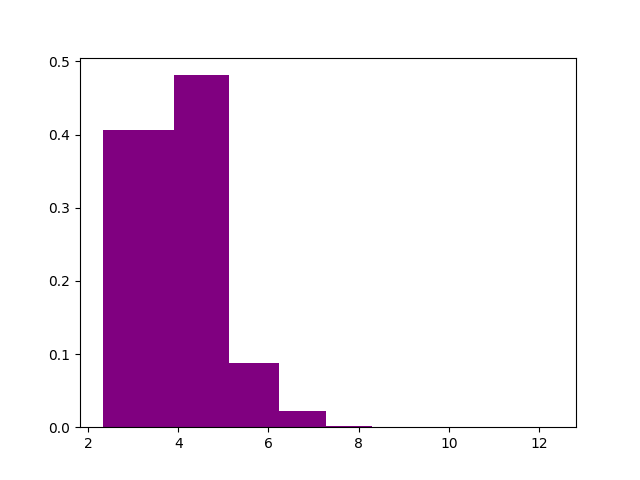}
    \caption{}
    \label{HistSize800}
    \end{subfigure}
     \begin{subfigure}{0.4\textwidth}
    \centering
    \includegraphics[width=\linewidth]{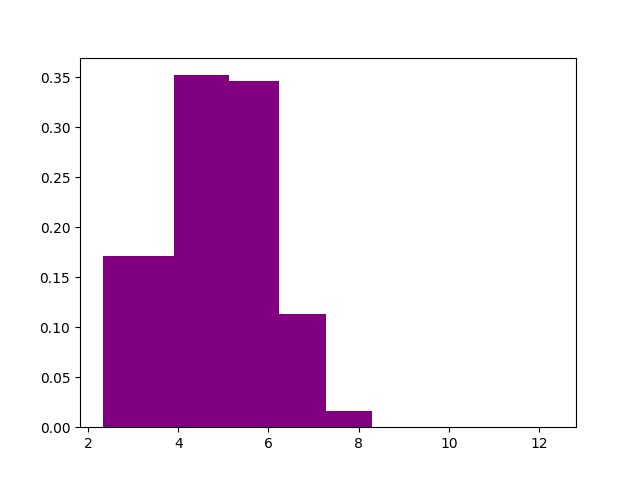}
    \caption{}
    \label{HistSize810}
    \end{subfigure}
    \begin{subfigure}{0.4\textwidth}
    \centering
    \includegraphics[width=\linewidth]{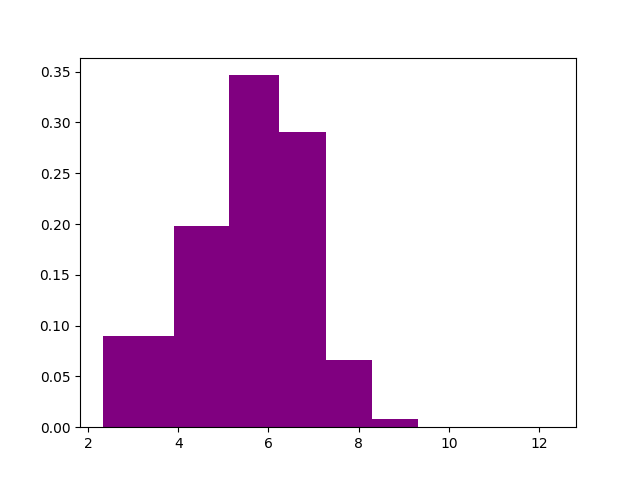}
    \caption{}
    \label{HistSize825}
    \end{subfigure}   
\caption{Normalized histograms of species abundances during the transition of $q_{min}$ from $0.8$ to $0.9$ for $M_0=1300$, $R=5.0$ and IP settings  at generation (a) 799, (b) 800, (c) 810 and (d) 825. The x-axis correspond to $\log_2(n)$, where $n$ is species size. Abundance histograms were constructed by using 10 replicas of the simulation.}
\label{HistSize}
\end{figure}

Another way to probe the effects of increase in $q_{min}$ is by computing species abundance distributions across the transition. Fig. \ref{HistSize} shows normalized abundance histograms constructed by using 10 replicas of the simulation. The bins in the x-axis (species size) are computed in log scale as follows: we first define bin sizes by setting $x(1) = x_0$ and $x(k+1) = x(k) + 2^k \Delta$ and collect the number of species $n(k)$ with sizes between $x(k+1)$ and $x(k)$. The histogram is plotted as $n(k) \times \log_2(x(k))$ for $x_0=\Delta=5$. Before the transition, at generation $t=799$, the histogram has the typical log-normal shape \cite{de2009global}. Right at the transition large species brake into several small species, causing the distribution to peak at low abundances. As the system overcomes the transient behavior, most of the small species become extinct while a few grow in size and become dominant again, as shown in generations $t=810$ and $825$. The abundance distribution eventually reaches a new equilibrium and recovers the log-normal shape. 

\begin{figure}[h]
     \begin{subfigure}{0.4\textwidth}
    \centering
    \includegraphics[width=\linewidth]{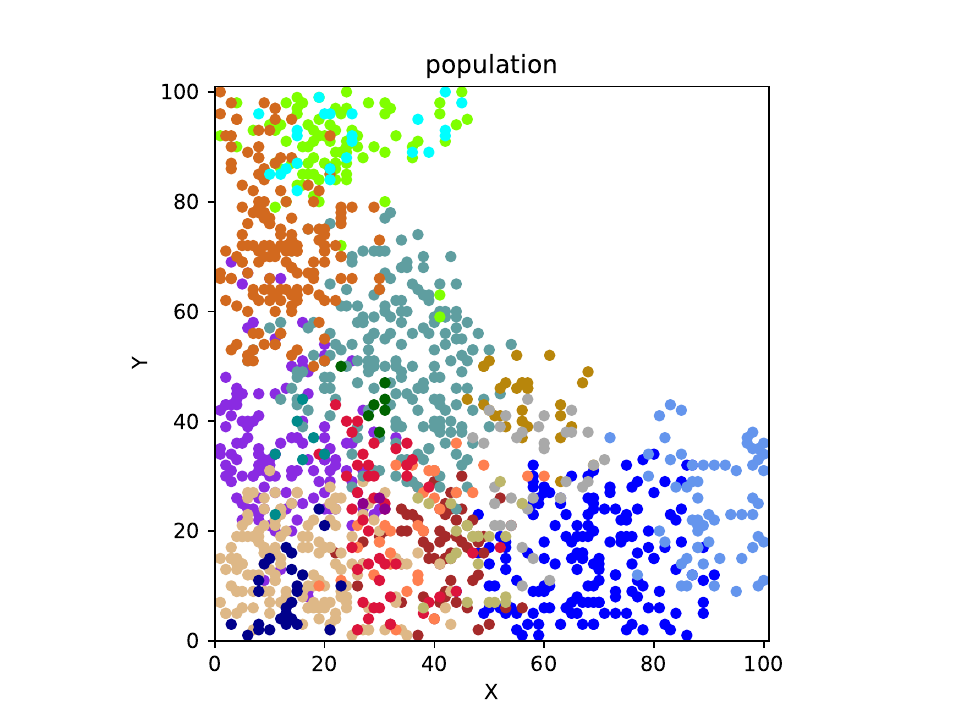} 
    \caption{}
    \label{SpDist_ff}
    \end{subfigure}
    \begin{subfigure}{0.4\textwidth}
    \centering
        \includegraphics[width=\linewidth]{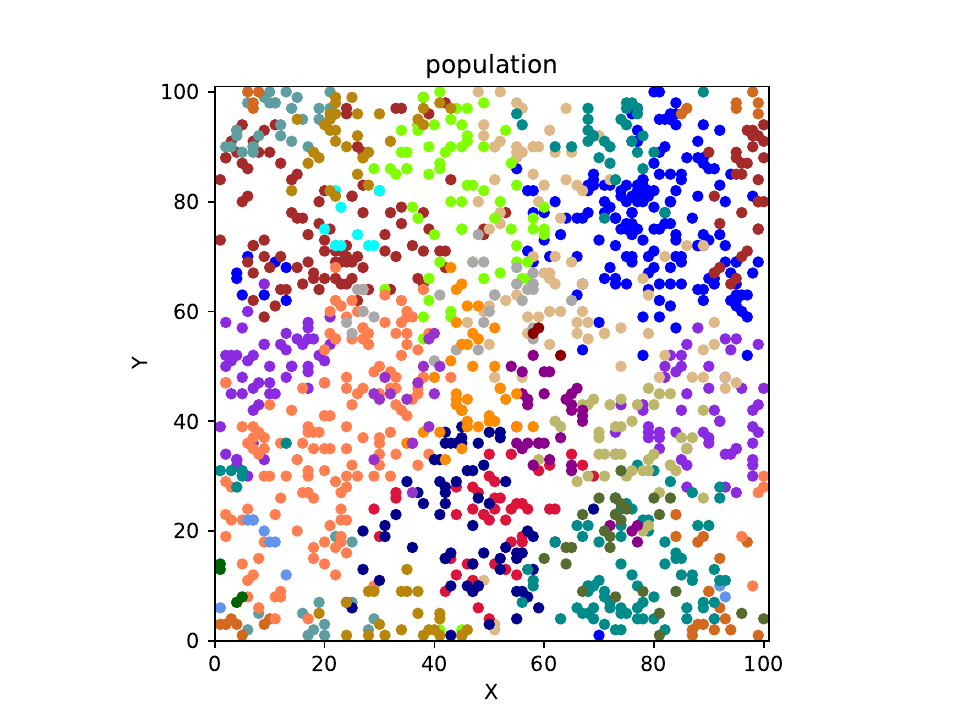}
    \caption{}
    \label{SpDist_fp}
    \end{subfigure}
     \begin{subfigure}{0.4\textwidth}
    \centering
    \includegraphics[width=\linewidth]{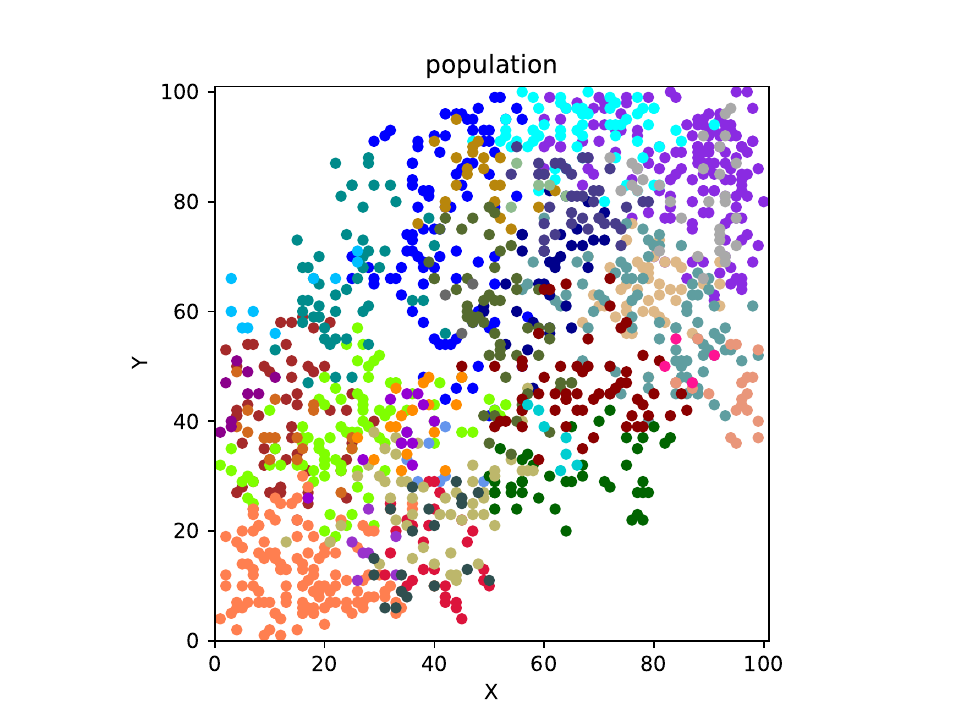}
    \caption{}
    \label{SpDist_if}
    \end{subfigure}
    \begin{subfigure}{0.4\textwidth}
    \centering
    \includegraphics[width=\linewidth]{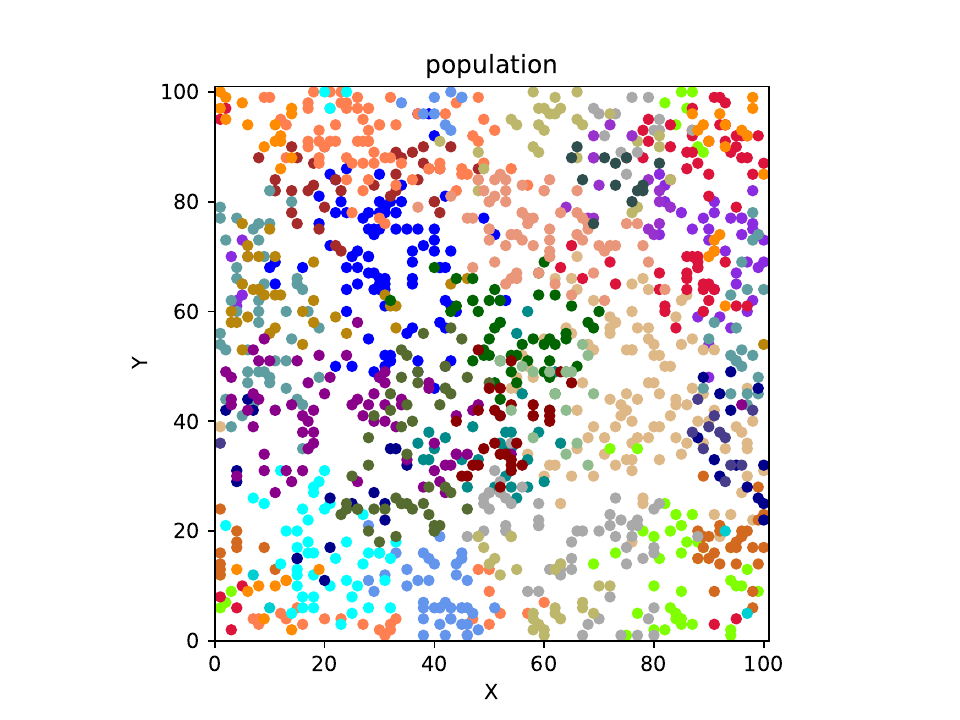}
    \caption{}
    \label{SpDist_ip}
    \end{subfigure}   
\caption{Spatial distribution of individual after 1500 generations, for $M_0=1300$ and $R=5.0$. In these simulations $q_{min}$ changed from $0.8$ to $0.9$ at $t=800$. (a) (FR) finite genome and reflective boundaries; (b) (FP) finite genome and periodic boundaries; (c) (IR) infinite genome and reflective boundaries and; (d) (IP) infinite genome and periodic boundaries. Species are represented by different colors.}
\label{SpDist}
\end{figure}

Before exploring variation of other sets of parameters during the evolutionary history, we show how the choice of boundaries, reflective or periodic, affects the spatial distribution of individuals. When the size of the mating neighborhood is much smaller than the living area, $R \ll L$, the physical and genetic spaces become strongly correlated, so that individuals of the same species tend to be located close together. However, the presence of reflective boundaries leads to non-homogeneous distribution of the population, that gets trapped in some corners whereas avoiding others. This is illustrated in Fig.\ref{SpDist}, showing the spatial distribution of all individuals at $T=1500$ for one particular realization of the dynamics. We can observe the clustering in both cases of reflective boundaries (Figs. \ref{SpDist_ff} and \ref{SpDist_if}), and the approximately homogeneous distribution along all space for periodic boundaries (Figs. \ref{SpDist_fp} and \ref{SpDist_ip}).

\begin{figure}[h]
    \begin{subfigure}{0.4\textwidth}
    \centering
    \includegraphics[width=\linewidth]{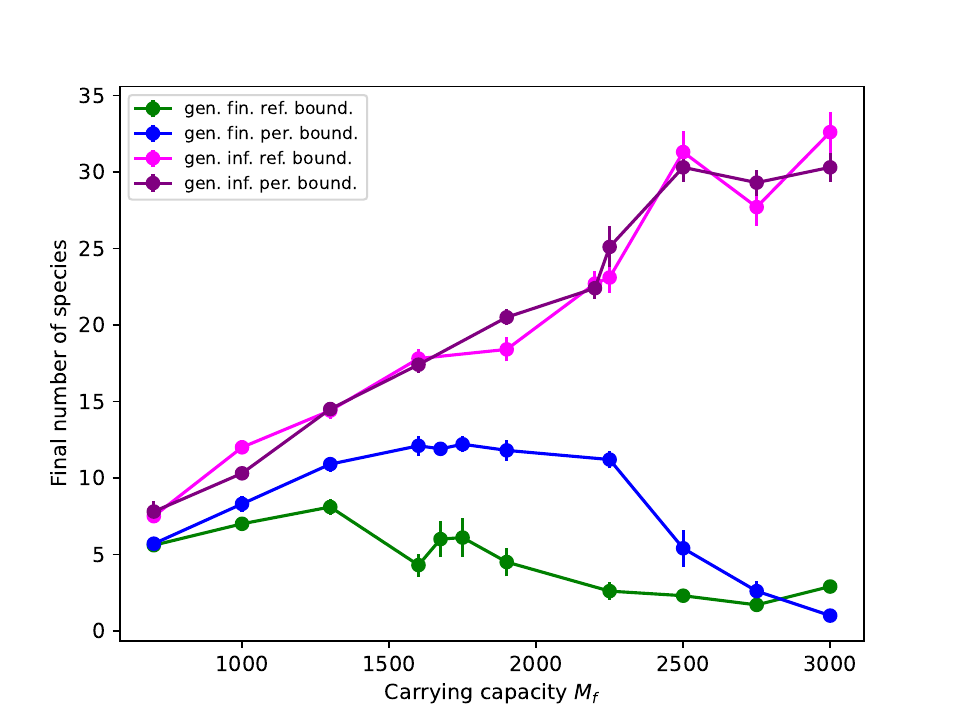}
    \caption{}
    \label{FinalNspc_Ntran}
    \end{subfigure}
    \begin{subfigure}{0.4\textwidth}
    \centering
    \includegraphics[width=\linewidth]{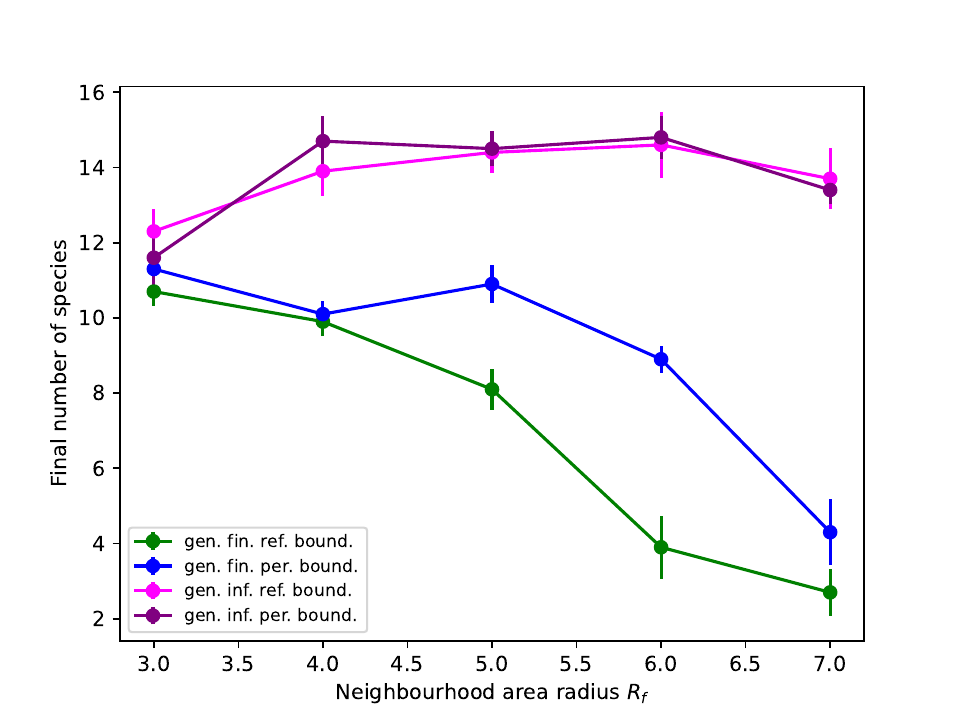}
    \caption{}
    \label{FinalNspc_Strans}
    \end{subfigure}   
\caption{Final number of species, at $T=1500$, when $q_{min}=0.8$. (a) initial carrying capacity $M_0=1300$ changed at $t=800$ to $M_f$ with $R=5$; (b) initial mating radius $R=5$ changed at $t=800$ to $R_f$ with $M_0=1300$. Results correspond to averages over 10 replicas of the simulations.}
\label{FinalNspc_trans}
\end{figure}

Variations in carrying capacity or mating neighborhood radius along the evolutionary history do not lead to drastic changes in diversity patterns, as populations adapt more easily to these new conditions. Nevertheless, the variation in the number of species shows very interesting and non-trivial patterns. Fig. \ref{FinalNspc_Ntran} shows that changes in carrying capacity from $M_0=1300$ to $M_f$ at $t=800$, despite changing the total number of individuals to values close to $M_f$, leads to a final number of species that depends critically on genome size. For  infinite genomes the number of species always increases with $M_f$, increasing diversity. For finite genomes, on the other hand, diversity increases until nearly $M_f=2200$ then promptly decreases reaching a single species for sufficiently large $M_f$. In this case, carrying capacity promotes the loss of diversity, leading to few species that dominate the environment. We note that, since $L$ is kept constant, changes in $M$ correspond to changes in population density. Corresponding changes in the mating neighborhood radius (starting from values large enough to admit compatible partners) also produces different behaviour according to genome size. The number of species stays nearly constant for infinite genomes, but decreases for finite ones, as can be seen in Fig. \ref{FinalNspc_Strans}.

\begin{figure}[h]
    \begin{subfigure}{0.4\textwidth}
    \centering
    \includegraphics[width=\linewidth]{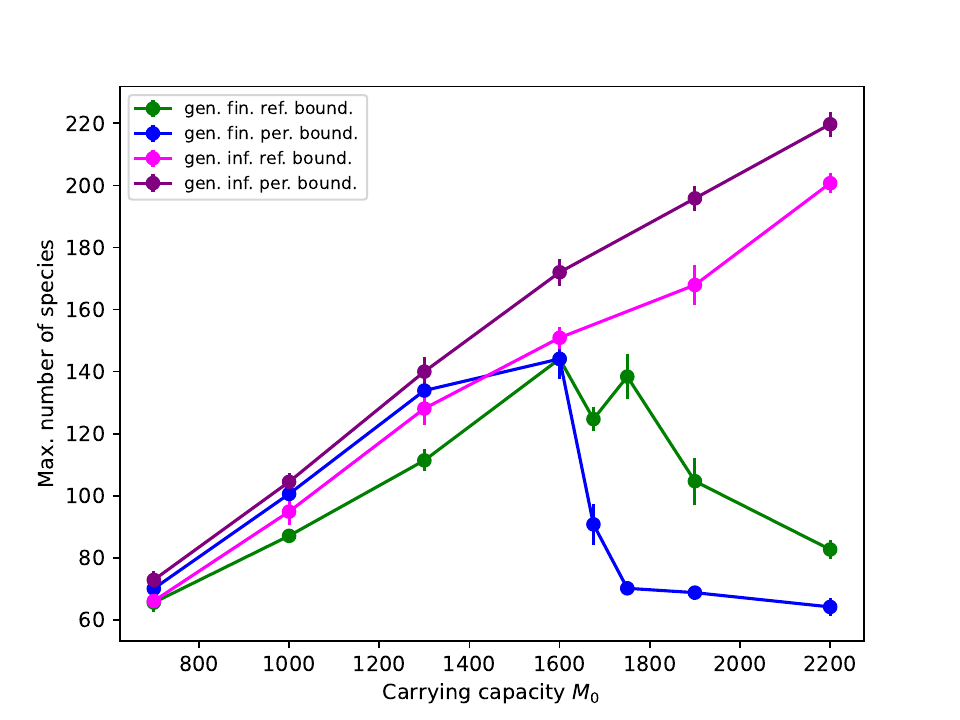}
    \caption{}
    \label{peakNspc_Nfix}
    \end{subfigure}
    \begin{subfigure}{0.4\textwidth}
    \centering
    \includegraphics[width=\linewidth]{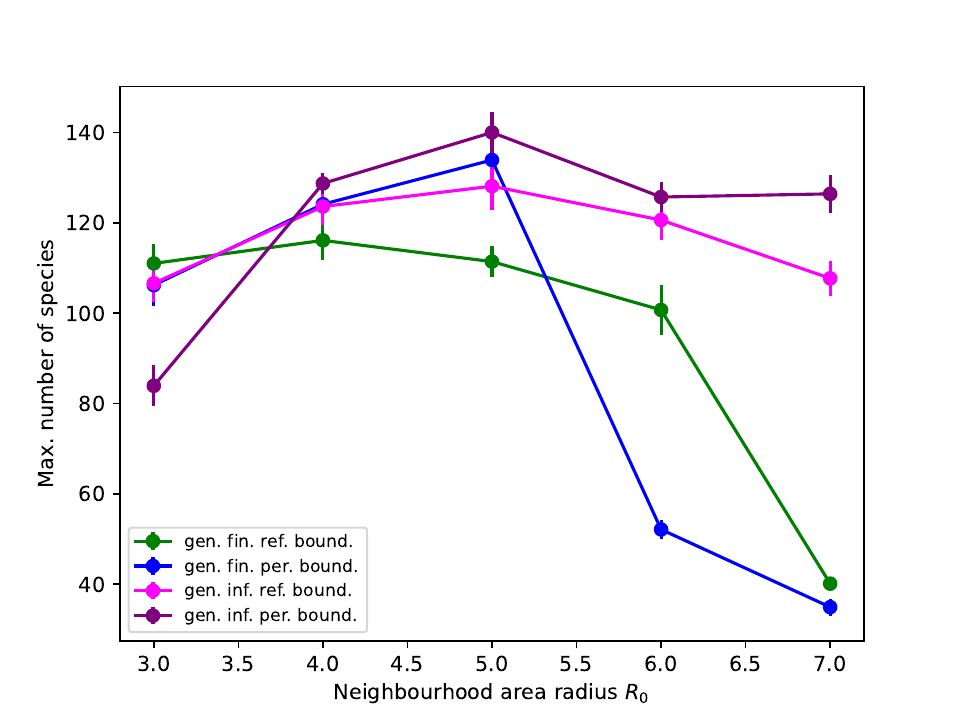}
    \caption{}
    \label{peakNspc_Sfix}
    \end{subfigure}   
\caption{Maximum number of species, at $t=800$, when $q_{min}$ changed from $0.8$ to $0.9$, for different choices of (a) carrying capacity $M_0$ with $R=5$; (b) mating radius $R$ with $M_0=1300$. Results correspond to averages over 10 replicas of the simulations.}
\label{peakNspc_fix}
\end{figure}

The values of $M_0$ and $R$ also matter when $q_{min}$ changes during evolution. To illustrate their role we compute the number of species at the peak of species explosion, as in Figs. \ref{Nspcgen} and \ref{Peaks}, as a function of these two parameters. Fig. \ref{peakNspc_Nfix} shows the number of species at $t=800$, when $q_{min}$ changed from 0.8 to 0.9, as a function of $M_0$ for $R=5$. We see that, while the maximum number of species increases linearly with $M_0$ for infinite genomes, the behavior for finite genomes is rather different. In this case, the number of species increases with $M_0$ up to a threshold value, where it starts to decrease. Moreover, the effect of boundary conditions is different in each case. For infinite genomes, boundary conditions do not change the number of species significantly, but for finite genomes it generally leads to more species.

The effect of $R$ is shown in Fig. \ref{peakNspc_Sfix}. For infinite genomes, the size of the mating neighborhood does not change the number of species too much, but for finite genomes $N_{spc}^{peak}$ drops as $R$ increases. Intuitively, larger $R$ means more gene flow and, therefore, less possibility of speciation. Interestingly, for infinite genomes this effect is compensated by the very larger number of loci where divergence can happen. 

\begin{figure}[h]
    \begin{subfigure}{0.4\textwidth}
    \centering
    \includegraphics[width=\linewidth]{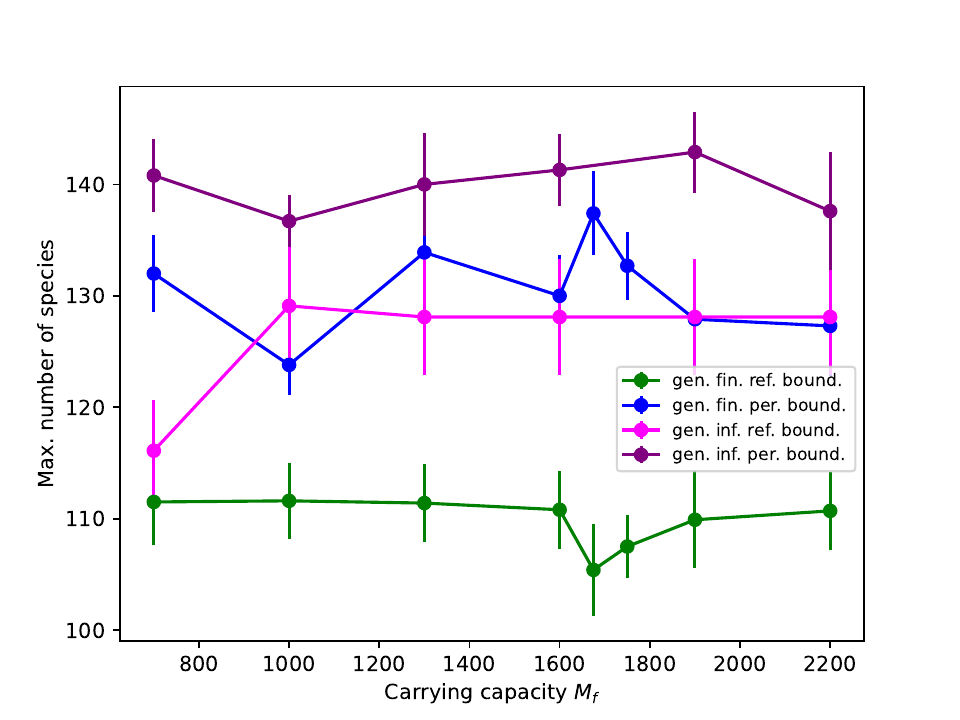}
    \caption{}
    \label{peakNspc_Ntran}
    \end{subfigure}
    \begin{subfigure}{0.4\textwidth}
    \centering
    \includegraphics[width=\linewidth]{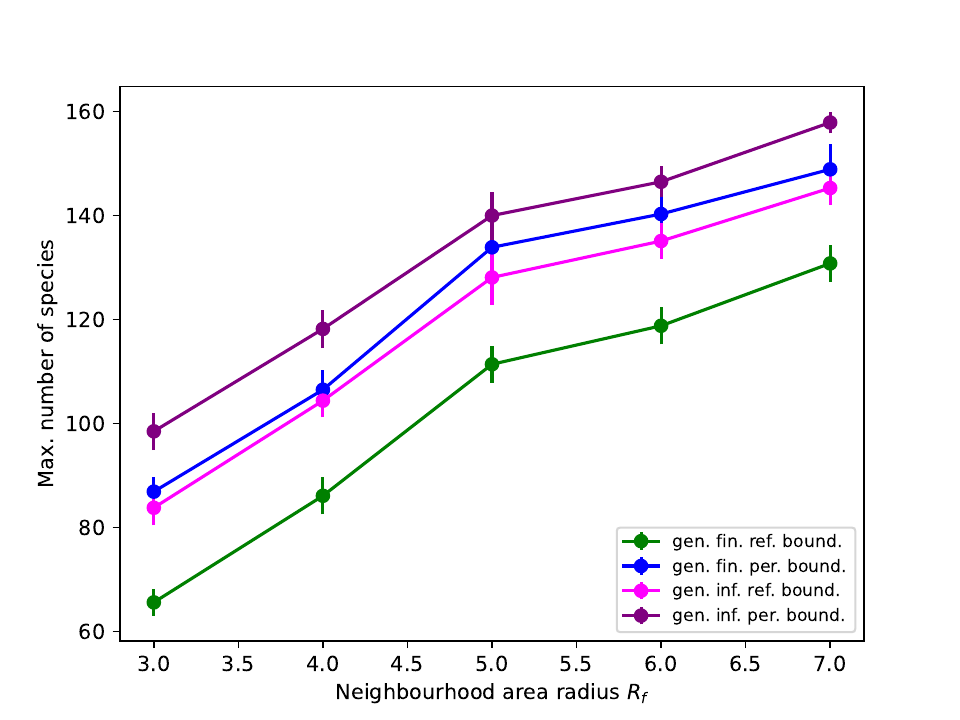}
    \caption{}
    \label{peakNspc_Strans}
    \end{subfigure}   
\caption{Maximum number of species, at $t=800$, when $q_{min}$ changed from $0.8$ to $0.9$. (a) initial carrying capacity $M_0=1300$ changed simultaneously at $t=800$ to $M_f$ with $R=5$; (b) initial mating radius $R=5$ changed simultaneously at $t=800$ to $R_f$ with $M_0=1300$. Results correspond to averages over 10 replicas of the simulations.}
\label{peakNspc_trans}
\end{figure}

Finally, Fig.\ref{peakNspc_trans} shows the maximum number of species $N_{spc}^{peak}$, at $t=800$, when $q_{min}$ changed from $0.8$ to $0.9$ together with changes in the carrying capacity (panel (a)) or size of mating radius (panel (b)). Although the carrying capacity does not change the number of species significantly, $N_{spc}^{peak}$ increases linearly with $R$ for all genome sizes and boundary conditions.

\section{Discussion and Conclusions}
\label{sec::discussion}

Changes in the environment can have significant consequences on biodiversity. When changes are fast compared to evolutionary time scales, it makes sense to investigate its effects using neutral models. In this work we introduced an individual based model where control parameters can be modified during the evolution, simulating sudden environmental changes. 

Starting with a genetically identical and uniformly distributed population, we introduced changes in parameter values (pulse events) after $800$ generations. We found out that the system's response to these changes are immediate, leading to effects such as drops, increases or peaks in species diversity. The choice of parameters changing at the pulse are proxy for environmental changes that lead to different carrying capacities, mating ranges and intensity of assortative mating \cite{seehausen2008speciation, seehausen1997cichlid, candolin2019mate, jeltsch2011consequences, royer2025temperature}. Simulations were performed for reflective and periodic boundary conditions and for genome length $1500$ and infinity. 

\noindent {\it {\bf Similarity threshold:}} Among the parameters changed at $t=800$, only the similarity threshold ($q_{min}$) lead to a drastic signature of the event, namely, a pronounced peak in diversity. This signature, however, was transient, and after the ``explosion" of new and small (regarding the number of individuals) species, the peak is followed by an fast decrease, showing that  most of the species were short-lived. The sudden rise in the number of species occurs when $q_{min}$ is increased. In this case, pairs that were compatible become abruptly incompatible, breaking the existing species into much smaller groups of compatible individuals. As species with small abundances went extinct, the remaining ones became larger (more abundant) until a stationary distribution was reached. The whole process takes about $50$ generations, from the explosion to the new equilibrium. A similar signature is also seen in the number of individuals that die without leaving descendants, and in the number of mating pairs having more than one offspring. Therefore, alongside the appearance of new small species there is a burst of deaths followed by an ``explosion" of multiple offspring by remaining compatible pairs of parents. The size of the peak in diversity depends on the boundary conditions and on genome length. Interestingly, decreasing the similarity threshold (making compatibility criterion less stringent) leads to a sudden but moderate decrease in the number of species (no peaks or valleys are observed). These processes give rise to a type of hysteresis if $q_{min}$ first increases and than decreases back, as the processes are not symmetric.

\noindent {\it {\bf Carrying capacity and mating range:}} Although sudden changes in threshold similarity leads to interesting responses, their occurrence in natural systems are not so usual. Changes in water turbidity is one example \cite{seehausen2008speciation}. Environmental changes leading to higher or lower carrying capacity, on the other hand, are quite common, as these can be promoted by increase or decrease in rainfall or incidence of solar radiation, for example. In this case we have seen that, depending on the genome length associated with reproduction, an increase in carrying capacity might promote diversification (in the case of infinite length) or loss of diversity. A similar pattern was observed with respect to changes in mating range.

These results can be understood in the light of the sympatric theory developed by Higgs and Derrida \cite{higgs1991stochastic} and further studied in \cite{de2017speciation,marquioni2025transition,braha2026sets}. In the sympatric limit where $S \to L$ and the genome length becomes very large, $B \to \infty$, the condition for speciation is $q_{min} > q_0$, where $q_0 = (1+4\mu M)^{-1}$ \cite{higgs1991stochastic}. However, this condition fails if the genome length is too small \cite{de2017speciation}. For given population size $M$ and mutation rate $\mu$, there is a minimum genome length $B_c$ such that only for $B > B_c$ speciation occurs. Moreover, it was shown that when the combination $4\mu M = O(1)$ $B_c$ increases with population size as $M^{3/2}$ \cite{braha2026sets}. Although these results were only demonstrated for sympatric scenarios, we observe that they also hold in the present case of parapatry ($S < L$). For $M=1300$ the value $B=1500$ was large enough and speciation was observed. However, when population size increases, the minimum genome length $B_c$ becomes larger than 1500 and the equilibrium configuration collapses to a single species scenario. For infinite genome length, on the other hand, speciation was still possible. This also explains the dependence on peak size on the carrying capacity shown in Fig. \ref{peakNspc_fix}.

As a general conclusion, we observed that the effects of changes can have significant transient responses, although the asymptotic equilibrium state of the system always converges to the configuration dictated by the final parameters. Moreover, the response of system to change in parameters is not symmetric, as the transient dynamics is very different if parameters increase or decrease. Since the equilibrium configuration does not depend on the history, the process cannot be characterized as canonical hysteresis, although the transient response does depend on the direction of change.

\begin{acknowledgements}
This work was partly supported by FAPESP [Grant 2021/ 14335-0 (M.A.M.A.)], by CNPq, Brazil, [Grant 303814/2023-3 (M.A.M.A.)] and by CAPES [Grant 88887.999656/2024-00].
The authors would like to thank Dr. Flávia M. D. Marquitti for several suggestions that improved the clarity of the manuscript.
\end{acknowledgements}

\bibliographystyle{unsrt}

\begin{thebibliography}{10}
	
	\bibitem{coyne2004speciation}
	Jerry~A Coyne, H~Allen Orr, et~al.
	\newblock {\em Speciation}, volume~37.
	\newblock Sinauer associates Sunderland, MA, 2004.
	
	\bibitem{nosil2012ecological}
	Patrik Nosil.
	\newblock {\em Ecological speciation}.
	\newblock Oxford University Press, 2012.
	
	\bibitem{dornelas2023looking}
	Maria Dornelas, Jonathan~M Chase, Nicholas~J Gotelli, Anne~E Magurran, Brian~J
	McGill, Laura~H Ant{\~a}o, Shane~A Blowes, Gergana~N Daskalova, Brian Leung,
	In{\^e}s~S Martins, et~al.
	\newblock Looking back on biodiversity change: lessons for the road ahead.
	\newblock {\em Philosophical Transactions of the Royal Society B: Biological
		Sciences}, 378(1881):20220199, 2023.
	
	\bibitem{waldock2018temperature}
	Conor Waldock, Maria Dornelas, and Amanda~E Bates.
	\newblock Temperature-driven biodiversity change: disentangling space and time.
	\newblock {\em Bioscience}, 68(11):873--884, 2018.
	
	\bibitem{hagen2012biodiversity}
	Melanie Hagen, W~Daniel Kissling, Claus Rasmussen, Marcus~AM De~Aguiar, Lee~E
	Brown, Daniel~W Carstensen, Isabel Alves-Dos-Santos, Yoko~L Dupont,
	Francois~K Edwards, Julieta Genini, et~al.
	\newblock Biodiversity, species interactions and ecological networks in a
	fragmented world.
	\newblock In {\em Advances in ecological research}, volume~46, pages 89--210.
	Elsevier, 2012.
	
	\bibitem{benton2016origins}
	Michael~J Benton.
	\newblock Origins of biodiversity.
	\newblock {\em PLoS biology}, 14(11):e2000724, 2016.
	
	\bibitem{thomas2004extinction}
	Chris~D Thomas, Alison Cameron, Rhys~E Green, Michel Bakkenes, Linda~J
	Beaumont, Yvonne~C Collingham, Barend~FN Erasmus, Marinez~Ferreira
	De~Siqueira, Alan Grainger, Lee Hannah, et~al.
	\newblock Extinction risk from climate change.
	\newblock {\em Nature}, 427(6970):145--148, 2004.
	
	\bibitem{urban2015accelerating}
	Mark~C Urban.
	\newblock Accelerating extinction risk from climate change.
	\newblock {\em Science}, 348(6234):571--573, 2015.
	
	\bibitem{maclean2011recent}
	Ilya~MD Maclean and Robert~J Wilson.
	\newblock Recent ecological responses to climate change support predictions of
	high extinction risk.
	\newblock {\em Proceedings of the National Academy of Sciences},
	108(30):12337--12342, 2011.
	
	\bibitem{bellard2012impacts}
	C{\'e}line Bellard, Cleo Bertelsmeier, Paul Leadley, Wilfried Thuiller, and
	Franck Courchamp.
	\newblock Impacts of climate change on the future of biodiversity.
	\newblock {\em Ecology letters}, 15(4):365--377, 2012.
	
	\bibitem{vrba1985environment}
	Elisabeth~S Vrba.
	\newblock Environment and evolution: alternative causes of the temporal
	distribution of evolutionary events.
	\newblock {\em South African Journal of Science}, 81(5):229--236, 1985.
	
	\bibitem{vrba1993turnover}
	Elisabeth~S Vrba.
	\newblock Turnover-pulses, the red queen, and related topics.
	\newblock {\em American Journal of Science}, 293(A):418--452, 1993.
	
	\bibitem{desbois2025elisabeth}
	Dominique Desbois et~al.
	\newblock Elisabeth vrba (1942-2025): s{\'e}lection et sp{\'e}ciation en
	r{\'e}ponse aux changements environnementaux.
	\newblock 2025.
	
	\bibitem{de2009global}
	Marcus A~M de~Aguiar, Michel Baranger, EM~Baptestini, L~Kaufman, and Y~Bar-Yam.
	\newblock Global patterns of speciation and diversity.
	\newblock {\em Nature}, 460(7253):384--387, 2009.
	
	\bibitem{nelson2024neutral}
	Erik~D Nelson.
	\newblock Neutral speciation in realistic populations.
	\newblock {\em Theoretical Ecology}, 17(3):281--288, 2024.
	
	\bibitem{baptestini2013role}
	Elizabeth~M Baptestini, Marcus~AM de~Aguiar, and Yaneer Bar-Yam.
	\newblock The role of sex separation in neutral speciation.
	\newblock {\em Theoretical ecology}, 6(2):213--223, 2013.
	
	\bibitem{martins2013evolution}
	Ayana~B Martins, Marcus~AM de~Aguiar, and Yaneer Bar-Yam.
	\newblock Evolution and stability of ring species.
	\newblock {\em Proceedings of the National Academy of Sciences},
	110(13):5080--5084, 2013.
	
	\bibitem{schneider2014toward}
	David~M Schneider, Eduardo do~Carmo, Ayana~B Martins, and Marcus~AM de~Aguiar.
	\newblock Toward a theory of topopatric speciation: the role of genetic
	assortative mating.
	\newblock {\em Physica A: Statistical Mechanics and its Applications},
	409:35--47, 2014.
	
	\bibitem{schneider2016diploid}
	David~M Schneider, Elizabeth~M Baptestini, and Marcus~AM de~Aguiar.
	\newblock Diploid versus haploid models of neutral speciation.
	\newblock {\em Journal of biological physics}, 42(2):235--245, 2016.
	
	\bibitem{de2017speciation}
	Marcus~AM de~Aguiar.
	\newblock Speciation in the derrida--higgs model with finite genomes and
	spatial populations.
	\newblock {\em Journal of Physics A: Mathematical and Theoretical},
	50(8):085602, 2017.
	
	\bibitem{marquioni2025transition}
	Vitor~M Marquioni and Marcus~AM de~Aguiar.
	\newblock The transition to speciation in the finite genome derrida--higgs
	model: a heuristic solution.
	\newblock {\em Journal of Physics A: Mathematical and Theoretical},
	58(17):175601, 2025.
	
	\bibitem{schneider2016mutation}
	David~M Schneider, Ayana~B Martins, and Marcus~AM de~Aguiar.
	\newblock The mutation--drift balance in spatially structured populations.
	\newblock {\em Journal of theoretical biology}, 402:9--17, 2016.
	
	\bibitem{baptestini2013conditions}
	Elizabeth~M Baptestini, Marcus~AM de~Aguiar, and Yaneer Bar-Yam.
	\newblock Conditions for neutral speciation via isolation by distance.
	\newblock {\em Journal of theoretical biology}, 335:51--56, 2013.
	
	
	\bibitem{oliveira2014modelo}
	Sergio Candido~de Oliveira~Junior.
	\newblock {\em Modelo baseado em agentes para especia{\c{c}}{\~a}o
		topop{\'a}trica}.
	\newblock PhD thesis, Universidade de S{\~a}o Paulo, 2014.
	
	\bibitem{botelho2022extinction}
	Larissa~Lubiana Botelho, Flavia Maria~Darcie Marquitti, and Marcus~AM
	de~Aguiar.
	\newblock Extinction and hybridization in a neutral model of speciation.
	\newblock {\em Journal of Physics A: Mathematical and Theoretical},
	55(38):385601, 2022.
	
	\bibitem{schneider2014topopatric}
	David~M Schneider.
	\newblock Topopatric speciation: From simulations to theory.
	\newblock In {\em Evolutionary Biology: Genome Evolution, Speciation,
		Coevolution and Origin of Life}, pages 357--367. Springer, 2014.
	
	\bibitem{princepe2022diversity}
	D{\'e}bora Princepe, Simone Czarnobai, Thiago~M Pradella, Rodrigo~A Caetano,
	Flavia~MD Marquitti, Marcus~AM de~Aguiar, and Sabrina~BL Araujo.
	\newblock Diversity patterns and speciation processes in a two-island system
	with continuous migration.
	\newblock {\em Evolution}, 76(10):2260--2271, 2022.
	
	\bibitem{princepe2022mito}
	D{\'e}bora Princepe, Marcus~AM de~Aguiar, and Joshua~B Plotkin.
	\newblock Mito-nuclear selection induces a trade-off between species ecological
	dominance and evolutionary lifespan.
	\newblock {\em Nature Ecology \& Evolution}, 6(12):1992--2002, 2022.
	
	\bibitem{jentsch2019theory}
	Anke Jentsch and Peter White.
	\newblock A theory of pulse dynamics and disturbance in ecology.
	\newblock {\em Ecology}, 100(7):e02734, 2019.
	
	\bibitem{martire2015carrying}
	Salvatore Martire, Valentina Castellani, and Serenella Sala.
	\newblock Carrying capacity assessment of forest resources: Enhancing
	environmental sustainability in energy production at local scale.
	\newblock {\em Resources, Conservation and Recycling}, 94:11--20, 2015.
	
	\bibitem{seehausen1997cichlid}
	Ole Seehausen, Jacques JM~van Alphen, and Frans Witte.
	\newblock Cichlid fish diversity threatened by eutrophication that curbs sexual
	selection.
	\newblock {\em Science}, 277(5333):1808--1811, 1997.
	
	\bibitem{seehausen2008speciation}
	Ole Seehausen, Yohey Terai, Isabel~S Magalhaes, Karen~L Carleton, Hillary~DJ
	Mrosso, Ryutaro Miyagi, Inke Van Der~Sluijs, Maria~V Schneider, Martine~E
	Maan, Hidenori Tachida, et~al.
	\newblock Speciation through sensory drive in cichlid fish.
	\newblock {\em Nature}, 455(7213):620--626, 2008.
	
	\bibitem{lizarraga2024assortativity}
	Joao~UF Liz{\'a}rraga, Flavia~MD Marquitti, and Marcus~AM de~Aguiar.
	\newblock Assortativity in sympatric speciation and species classification.
	\newblock {\em Physica A: Statistical Mechanics and its Applications},
	653:130111, 2024.
	
	\bibitem{higgs1991stochastic}
	Paul~G Higgs and Bernard Derrida.
	\newblock Stochastic models for species formation in evolving populations.
	\newblock {\em Journal of Physics A: Mathematical and General}, 24(17):L985,
	1991.
	
	\bibitem{candolin2019mate}
	Ulrika Candolin and Bob~BM Wong.
	\newblock Mate choice in a polluted world: consequences for individuals,
	populations and communities.
	\newblock {\em Philosophical Transactions of the Royal Society B: Biological
		Sciences}, 374(1781):20180055, 2019.
	
	\bibitem{jeltsch2011consequences}
	Florian Jeltsch, Kirk~A Moloney, Monika Schwager, Katrin K{\"o}rner, and Niels
	Blaum.
	\newblock Consequences of correlations between habitat modifications and
	negative impact of climate change for regional species survival.
	\newblock {\em Agriculture, ecosystems \& environment}, 145(1):49--58, 2011.
	
	\bibitem{royer2025temperature}
	Lucie Royer and Jacques R{\'e}gni{\`e}re.
	\newblock Temperature and morphology affect the performance and cost of flight
	in spruce budworm females.
	\newblock {\em Ecology and Evolution}, 15(12):e72529, 2025.
	
	\bibitem{braha2026sets}
	Dan Braha, Marcus~AM de~Aguiar, and Vitor~M Marquioni.
	\newblock What sets the critical genome length for sympatric speciation? a
	closed form and asymptotic theory.
	\newblock {\em arXiv preprint arXiv:2608.25995}, 2026.
	
\end{thebibliography}

\end{document}